\documentclass[12pt]{iopart}

\usepackage{graphicx,amssymb}
\usepackage{lineno}

\def\be{\begin{equation}}
\def\ee{\end{equation}}

\begin{document}


\title[Fermi-LAT Collaboration: Dark Matter Annihilation in Clusters of Galaxies]{Constraints on Dark Matter Annihilation in Clusters of Galaxies with the Fermi Large Area Telescope}

\noindent
M.~Ackermann$^{2}$, 
M.~Ajello$^{2}$, 
A.~Allafort$^{2}$, 
L.~Baldini$^{3}$, 
J.~Ballet$^{4}$, 
G.~Barbiellini$^{5,6}$, 
D.~Bastieri$^{7,8}$, 
K.~Bechtol$^{2}$, 
R.~Bellazzini$^{3}$, 
R.~D.~Blandford$^{2}$
% ...
\medskip
\begin{enumerate}
\item[1.] Corresponding authors: T.~E.~Jeltema, tesla@ucolick.org; S.~Profumo, profumo@scipp.ucsc.edu
\item[2.] W. W. Hansen Experimental Physics Laboratory, Kavli Institute for Particle Astrophysics and Cosmology, Department of Physics and SLAC National Accelerator Laboratory, Stanford University, Stanford, CA 94305, USA
\item[3.] Istituto Nazionale di Fisica Nucleare, Sezione di Pisa, I-56127 Pisa, Italy
\item[4.] Laboratoire AIM, CEA-IRFU/CNRS/Universit\'e Paris Diderot, Service d'Astrophysique, CEA Saclay, 91191 Gif sur Yvette, France
\item[5.] Istituto Nazionale di Fisica Nucleare, Sezione di Trieste, I-34127 Trieste, Italy
\item[6.] Dipartimento di Fisica, Universit\`a di Trieste, I-34127 Trieste, Italy
\item[7.] Istituto Nazionale di Fisica Nucleare, Sezione di Padova, I-35131 Padova, Italy
\item[8.] Dipartimento di Fisica ``G. Galilei", Universit\`a di Padova, I-35131 Padova, Italy
\item[9.] Istituto Nazionale di Fisica Nucleare, Sezione di Perugia, I-06123 Perugia, Italy
\item[10.] Dipartimento di Fisica, Universit\`a degli Studi di Perugia, I-06123 Perugia, Italy
\item[11.] Centre d'\'Etude Spatiale des Rayonnements, CNRS/UPS, BP 44346, F-30128 Toulouse Cedex 4, France
\item[12.] Department of Physics, Center for Cosmology and Astro-Particle Physics, The Ohio State University, Columbus, OH 43210, USA
\item[13.] Dipartimento di Fisica ``M. Merlin" dell'Universit\`a e del Politecnico di Bari, I-70126 Bari, Italy
\item[14.] Istituto Nazionale di Fisica Nucleare, Sezione di Bari, 70126 Bari, Italy
\item[15.] Laboratoire Leprince-Ringuet, \'Ecole polytechnique, CNRS/IN2P3, Palaiseau, France
\item[16.] Institut de Ciencies de l'Espai (IEEC-CSIC), Campus UAB, 08193 Barcelona, Spain
\item[17.] INAF-Istituto di Astrofisica Spaziale e Fisica Cosmica, I-20133 Milano, Italy
\item[18.] Space Science Division, Naval Research Laboratory, Washington, DC 20375, USA
\item[19.] George Mason University, Fairfax, VA 22030, USA
\item[20.] National Research Council Research Associate, National Academy of Sciences, Washington, DC 20001, USA
\item[21.] Laboratoire de Physique Th\'eorique et Astroparticules, Universit\'e Montpellier 2, CNRS/IN2P3, Montpellier, France
\item[22.] Department of Physics and Astronomy, Sonoma State University, Rohnert Park, CA 94928-3609, USA
\item[23.] Department of Physics, Stockholm University, AlbaNova, SE-106 91 Stockholm, Sweden
\item[24.] The Oskar Klein Centre for Cosmoparticle Physics, AlbaNova, SE-106 91 Stockholm, Sweden
\item[25.] Royal Swedish Academy of Sciences Research Fellow, funded by a grant from the K. A. Wallenberg Foundation
\item[26.] Dipartimento di Fisica, Universit\`a di Udine and Istituto Nazionale di Fisica Nucleare, Sezione di Trieste, Gruppo Collegato di Udine, I-33100 Udine, Italy
\item[27.] CNRS/IN2P3, Centre d'\'Etudes Nucl\'eaires Bordeaux Gradignan, UMR 5797, Gradignan, 33175, France
\item[28.] Universit\'e de Bordeaux, Centre d'\'Etudes Nucl\'eaires Bordeaux Gradignan, UMR 5797, Gradignan, 33175, France
\item[29.] Osservatorio Astronomico di Trieste, Istituto Nazionale di Astrofisica, I-34143 Trieste, Italy
\item[30.] Department of Physical Sciences, Hiroshima University, Higashi-Hiroshima, Hiroshima 739-8526, Japan
\item[31.] Agenzia Spaziale Italiana (ASI) Science Data Center, I-00044 Frascati (Roma), Italy
\item[32.] NASA Goddard Space Flight Center, Greenbelt, MD 20771, USA
\item[33.] Center for Space Plasma and Aeronomic Research (CSPAR), University of Alabama in Huntsville, Huntsville, AL 35899, USA
\item[34.] Santa Cruz Institute for Particle Physics, Department of Physics and Department of Astronomy and Astrophysics, University of California at Santa Cruz, Santa Cruz, CA 95064, USA
\item[35.] Research Institute for Science and Engineering, Waseda University, 3-4-1, Okubo, Shinjuku, Tokyo, 169-8555 Japan
\item[36.] Center for Research and Exploration in Space Science and Technology (CRESST) and NASA Goddard Space Flight Center, Greenbelt, MD 20771, USA
\item[37.] Department of Physics and Department of Astronomy, University of Maryland, College Park, MD 20742, USA
\item[38.] Istituto Nazionale di Fisica Nucleare, Sezione di Roma ``Tor Vergata", I-00133 Roma, Italy
\item[39.] Department of Physics and Astronomy, University of Denver, Denver, CO 80208, USA
\item[40.] Institute of Space and Astronautical Science, JAXA, 3-1-1 Yoshinodai, Sagamihara, Kanagawa 229-8510, Japan
\item[41.] Hiroshima Astrophysical Science Center, Hiroshima University, Higashi-Hiroshima, Hiroshima 739-8526, Japan
\item[42.] Max-Planck Institut f\"ur extraterrestrische Physik, 85748 Garching, Germany
\item[43.] Department of Physics, University of Washington, Seattle, WA 98195-1560, USA
\item[44.] Space Sciences Division, NASA Ames Research Center, Moffett Field, CA 94035-1000, USA
\item[45.] NYCB Real-Time Computing Inc., Lattingtown, NY 11560-1025, USA
\item[46.] Department of Chemistry and Physics, Purdue University Calumet, Hammond, IN 46323-2094, USA
\item[47.] Partially supported by the International Doctorate on Astroparticle Physics (IDAPP) program
\item[48.] Instituci\'o Catalana de Recerca i Estudis Avan\c{c}ats (ICREA), Barcelona, Spain
\item[49.] Department of Physics and Center for Space Sciences and Technology, University of Maryland Baltimore County, Baltimore, MD 21250, USA
\item[50.] Dipartimento di Fisica, Universit\`a di Roma ``Tor Vergata", I-00133 Roma, Italy
\item[51.] Department of Physics, Royal Institute of Technology (KTH), AlbaNova, SE-106 91 Stockholm, Sweden
\item[52.] School of Pure and Applied Natural Sciences, University of Kalmar, SE-391 82 Kalmar, Sweden
\end{enumerate}

\begin{abstract}

\noindent Nearby clusters and groups of galaxies are potentially bright sources of high-energy gamma-ray emission resulting from the pair-annihilation of dark matter particles.  However, no significant gamma-ray emission has been detected so far from clusters in the first 11 months of observations with the Fermi Large Area Telescope.  We interpret this non-detection in terms of constraints on dark matter particle properties.  In particular for leptonic annihilation final states and particle masses greater than $\sim 200$ GeV, gamma-ray emission from inverse Compton scattering of CMB photons is expected to dominate the dark matter annihilation signal from clusters, and our gamma-ray limits exclude large regions of the parameter space that would give a good fit to the recent anomalous Pamela and Fermi-LAT electron-positron measurements.  We also present constraints on the annihilation of more standard dark matter candidates, such as the lightest neutralino of supersymmetric models.  The constraints are particularly strong when including the fact that clusters are known to contain substructure at least on galaxy scales, increasing the expected gamma-ray flux  by a factor of $\sim 5$ over a smooth-halo assumption.  We also explore the effect of uncertainties in cluster dark matter density profiles, finding a systematic uncertainty in the constraints of roughly a factor of two, but similar overall conclusions. In this work, we focus on deriving limits on dark matter models; a more general consideration of the Fermi-LAT data on clusters and clusters as gamma-ray sources is forthcoming.

\end{abstract}

\maketitle

\section{Introduction}

Clusters of galaxies are the most massive objects in the Universe that have had time to virialize by the present epoch, making nearby clusters attractive targets for searches for a signature from dark matter annihilation.  Clusters are more distant, but more massive than dwarf spheroidal galaxies, and like dwarf spheroidals, they are very dark matter dominated and typically lie at high galactic latitudes where the contamination from Galactic gamma-ray background emission is low.  A common feature of the stable particle products resulting from the annihilation of weakly interacting massive particles (WIMPs), prime candidates for dark matter, is a significant gamma-ray emission, though in addition dark matter annihilation is expected to yield a broad multiwavelength spectrum of emission \cite{Colafrancesco:2005ji}.  For a variety of particle models, the gamma-ray signal from the annihilation of dark matter in clusters was predicted to be potentially observable by the Fermi Gamma-Ray Space Telescope \cite{je09,Pinzke:2009cp}.

In proto-typical WIMP models, such as for neutralinos in the minimal supersymmetric extension to the Standard Model, gamma-ray emission results primarily from the decay of neutral pions produced as part of the annihilation products.  For WIMPs annihilating primarily into leptons, such as the dark matter models proposed to explain the Pamela positron excess \cite{pamela}, gamma-ray emission results instead dominantly from internal bremsstrahlung in the final state.  In addition, these models lead to a large population of energetic $e^+$ and $e^-$ among the annihilation products (hence the popularity as an explanation for the recent $e^+e^-$ measurements) which in turn lead to significant gamma-ray emission through the inverse Compton scattering (IC) of cosmic microwave background photons (CMB) or other photon fields in the inter-galactic medium.  The IC emission is particularly important in clusters which effectively confine cosmic ray electrons much longer than the time it takes them to lose energy through this mechanism. We will show that Fermi-LAT observations of clusters place some of the best constraints on the possible interpretation of the local $e^+e^-$ signal observed by Pamela, Fermi-LAT, and H.E.S.S. \cite{pamela,fermiepem,hessepem} as a signal from the annihilation of dark matter \cite{Grasso:2009ma,Bergstrom:2009fa,Meade:2009iu}.

Dark matter annihilation is not the only potential source of gamma-ray emission from clusters of galaxies.  Clusters are also expected to host a population of cosmic rays accelerated in merger and accretion shocks and by AGN throughout the cluster's history.  Collisions of relativistic protons with protons in the thermal intracluster medium would lead to gamma-ray emission through the decays of neutral pions produced in the collision.  In addition, diffuse radio emission observed in many clusters reveals the presence of relativistic electrons, and similar to the dark matter case, IC scattering of background radiation by these relativistic electrons could also lead to detectable gamma-ray emission.  The shape of the gamma-ray spectrum is generally expected to be different than that from dark matter annihilation, although sufficient statistics are required to distinguish between the two \cite{je09}.

Significant gamma-ray emission has not been detected from local clusters by Fermi-LAT in the first 11 months of observations \cite{keith}, aside from the detection of the central radio galaxies in a couple of clusters \cite{n1275, m87}.  The Fermi upper limits on gamma-rays from clusters place interesting limits on 
dark matter models.
In this paper, we investigate in detail the implications of the Fermi-LAT data in terms of limits on some of the fundamental properties of dark matter, such as the particle mass, pair-annihilation final state and pair-annihilation rate.  We use 11 months of Fermi-LAT survey mode data and self-consistently derive the flux upper limits for dark matter models based on the expected gamma-ray spectrum as a function of particle mass, annihilation final state, and emission mechanism (Sec.~\ref{sec:analysis}).  For specific models, we present the Fermi exclusion regions for the best-candidate clusters in the annihilation cross-section - mass plane (Sec.~\ref{sec:results}).  In this work, we focus specifically on clusters as potential sources of a dark matter signal.  The Fermi-LAT results for a much larger sample of clusters as well as a more general consideration of clusters as gamma-ray sources, including constraints on their cosmic ray content, will be presented in a forthcoming paper.

\section{Data Analysis}\label{sec:analysis}

\subsection{Cluster Sample}\label{sec:cat}

We selected our cluster sample from among the clusters predicted to have the brightest gamma-ray emission from dark matter annihilation based on the modeling described in \cite{je09}.  The predicted gamma-ray flux from dark matter annihilation $\Phi_\gamma$ is proportional to the dark matter density squared integrated along the line-of-sight and averaged on a given angular scale:
\begin{equation}\label{eq:jpsi}
\Phi_\gamma\propto J=\frac{1}{\Delta\Omega}\int {\rm d}\Omega \int_{\rm l.o.s.}\rho^2(l){\rm d}l(\psi).
\end{equation}
In turn, for a distant object whose spatial extent is smaller than the angular scale under consideration, one has a simple scaling with the cluster distance (even though in the present study we resort to the full treatment in the equation above):
\begin{equation}\label{eq:jpsi}
J\simeq\frac{1}{D^2}\int_{\rm Vol}\ r^2 \rho^2(r) {\rm d}r
\end{equation}
where ${\rm Vol}$ indicates the volume of the cluster, and $D$ is the cluster distance. In determining the brightest predicted clusters, we used the HIFLUGCS \cite{re02} catalog of the brightest X-ray clusters in the sky and X-ray determinations of the cluster masses.  Starting from the best candidate clusters, we removed from the sample the Perseus cluster since its central galaxy, NGC 1275, hosts a bright gamma-ray AGN \cite{n1275}, the Ophiuchus cluster, which lies near the Galactic Center, and the Norma cluster, which lies near the Galactic plane.  Our final sample of six clusters is listed in Table \ref{sample}.

\begin{table}
\caption{Clusters considered in the Fermi-LAT analysis.  Cluster positions are taken from \cite{re02}, and redshifts from the NASA/IPAC Extragalactic Database (http://nedwww.ipac.caltech.edu/). The Fermi-LAT data is analyzed within a $10^{\circ}$ radius region surrounding each cluster position.  Column 5 lists $J$, the integral along the line of sight of the cluster dark matter density squared, for an assumed NFW profile modeled as described in sec.~\ref{sec:dm} and including the smooth halo only (i.e.~neglecting the potential effect of substructure). $J$ is calculated within an radius of one degree (i.e. within a solid angle region $\Delta\Omega$ corresponding to one degree). Centaurus and NGC 4636 are cool core clusters in X-ray \cite{Chen:2007sz}. \label{sample}}
\begin{indented}
\item[]\begin{tabular}{lcccc}
\br
Cluster & RA & Dec. & $z$ & $J$ ($10^{17}$ GeV$^2$ cm$^{-5}$)\\
\mr
AWM 7 &43.6229 &41.5781 &0.0172 &$1.4^{+0.1}_{-0.1}$  \\
Fornax &54.6686 &-35.3103 &0.0046 &$6.8^{+1.0}_{-0.9}$  \\
M49 &187.4437 &7.9956 &0.0033 &$4.4^{+0.2}_{-0.1}$  \\
NGC 4636 &190.7084 &2.6880 &0.0031 &$4.1^{+0.3}_{-0.3}$  \\
Centaurus (A3526) &192.1995 &-41.3087 &0.0114 &$2.7^{+0.1}_{-0.1}$  \\
Coma &194.9468 &27.9388 &0.0231 &$1.7^{+0.1}_{-0.1}$  \\
\br
\end{tabular}
\end{indented}
\end{table}

\subsection{Fermi-LAT Observations and Reduction}\label{sec:data}

The Fermi-LAT data selection is the same as described in \cite{dsph}.  In brief, we utilize the first 11 months of scanning mode observations collected between August 4, 2008 and July 4, 2009.  Fermi operates in sky survey mode, covering the entire sky every 3 hours and providing nearly uniform coverage.  We employ the standard data cuts and background rejection defined in \cite{LAT} keeping only ``Diffuse'' class events which have a high probability of being gamma-rays.  In addition, to avoid contamination from the Earth's ``albedo" gamma-ray emission\footnote{gamma-ray emission from cosmic ray interactions with the Earth's atmosphere}, we exclude time intervals where the zenith angle was greater than 105$^\circ$ (the Earth's limb lies at a zenith angle of 113$^\circ$ for Fermi) and times when the rocking angle (angle between the viewing direction and the zenith) was greater than 43$^\circ$.  The data were further limited to include only events with energies between 100 MeV and 100 GeV, where the LAT instrument is best calibrated and where models of the diffuse gamma-ray backgrounds are available\footnote{http://fermi.gsfc.nasa.gov/ssc/data/analysis/LAT\_caveats.html}, and only data within a 10 degree radius of each cluster. 

The data analysis was carried out using the Fermi Science Tools package version 9.15.2 and using the Pass 6 version 3 LAT instrumental response functions (IRFs).  The LAT data were fit using an unbinned likelihood method \cite{ca79,ma96} as implemented in the pyLikelihood task in the Science Tools.  This task fits the LAT data to a combined spatial and spectral model of the source folded with the LAT response and point spread function (PSF) using a maximum likelihood statistic.  To account for the gamma-ray backgrounds, the fit includes models for both the Galactic and isotropic diffuse gamma-ray emission\footnote{http://fermi.gsfc.nasa.gov/ssc/data/access/lat/BackgroundModels.html} (gll\_iem\_v02.fit and isotropic\_iem\_v02.txt, respectively) as well as surrounding point sources within a 15 degree radius.  The diffuse emission models are those currently advocated and released by the Fermi-LAT collaboration, and the normalizations of these components are allowed to vary during the fitting.  The background point sources are drawn from the First Fermi-LAT source catalog, and the parameters of these sources are left fixed at their best-fit catalog values \cite{1fgl}\footnote{http://fermi.gsfc.nasa.gov/ssc/data/access/lat/1yr\_catalog/}.

No significant detections of gamma-ray emission from clusters of galaxies have been found in the first 11 months of Fermi-LAT observations \cite{keith} as will also be discussed in a forthcoming paper.  In general, the background model including the Galactic diffuse, isotropic diffuse, and nearby point sources gives a good fit to the data.  The most noticeable residual in the fits is the tendency for the model to slightly overpredict the data by around 10\% at the lowest energies considered (100-150 MeV), likely due to the rapidly changing effective area at these energies.  However, for the models discussed in this paper the flux upper limits we derive are essentially unaffected.  The improvement in the LAT sensitivity and PSF at higher energies and the softness of the diffuse backgrounds mean that the dark matter models we consider are primarily constrained by the LAT data at higher energies.

In order to place upper limits on the possible gamma-ray flux from dark matter annihilation, we model each cluster as a Fermi point source with an assumed dark matter annihilation spectrum.  In particular, we fit the data for a grid of assumed dark matter particle masses and a couple of representative annihilation final states.  
Upper limits on the flux are calculated using a profile likelihood technique: values of the flux are scanned fitting with respect to the other free parameters and the change in the log-likelihood is calculated.  The 95\% confidence upper limit on the flux is defined as the flux value for a change in the log-likelihood of 1.36.

\subsection{Flux upper limits for selected dark matter models}

We analyze the Fermi-LAT data to derive upper limits on the gamma-ray flux from the clusters of galaxies listed in Table 1.  The Fermi flux limits depend on the shape of the assumed gamma-ray spectrum, and we consider dark matter annihilation in specific representative models, self-consistently incorporating the expected spectral shape for a given particle mass and annihilation final state.  In Sec.~\ref{sec:results}, we will describe the translation of the gamma-ray flux limits to limits on the dark matter annihilation cross-section including modeling of the cluster dark matter density profiles.

\begin{figure}[t]
\begin{center}
\includegraphics[width=11cm,clip]{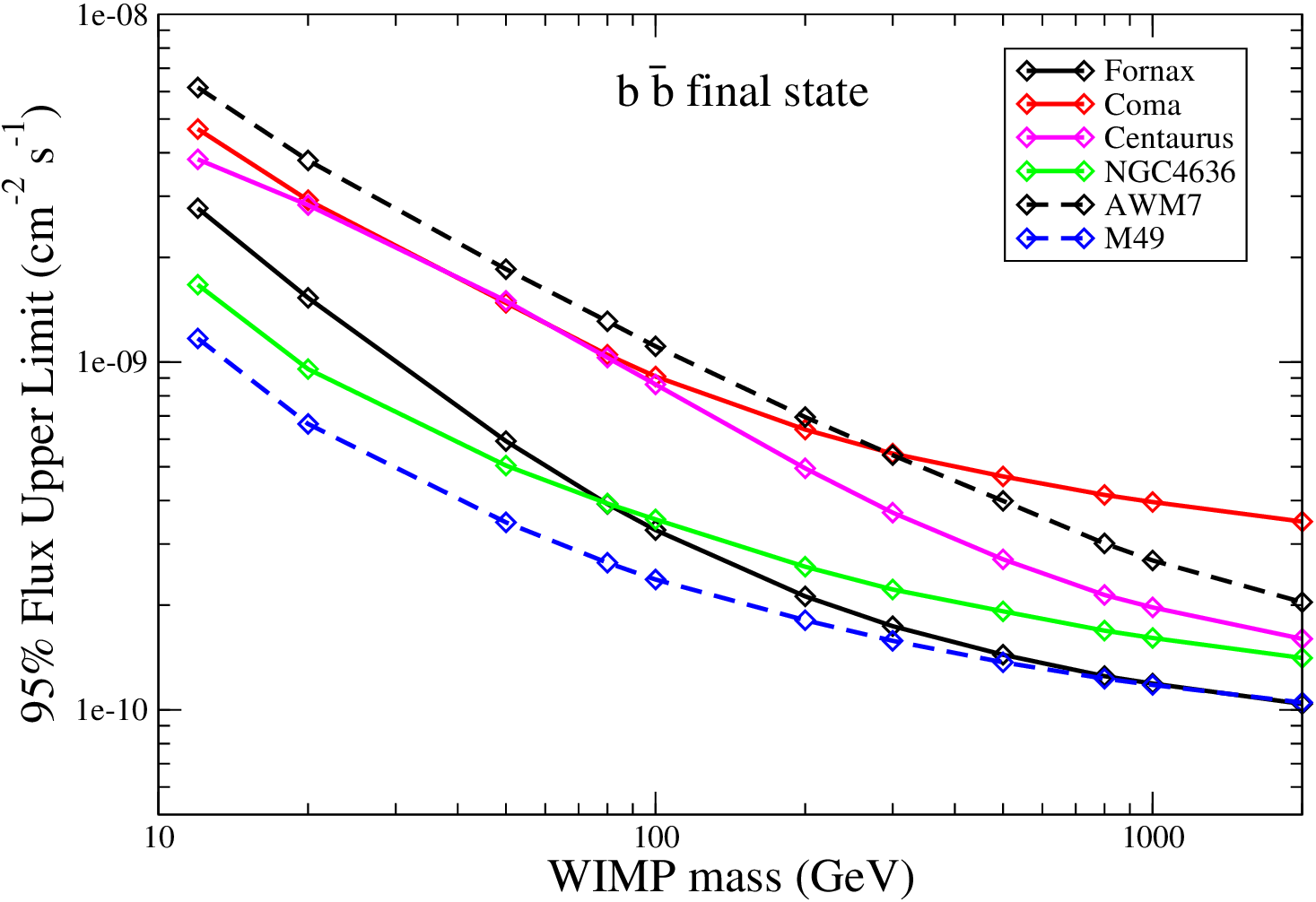}\\
\caption{Flux upper limits as a function of particle mass for an assumed $b\bar b$ final state for the clusters listed in Table \ref{sample}.  The upper limits reflect the 95\% confidence level limits on the photon flux in the 100 MeV to 100 GeV energy range. \label{fig:bblim}}
\end{center}
\end{figure}

Fig.~\ref{fig:bblim} shows the 95\% confidence upper limits on the flux in the 100 MeV to 100 GeV band for the gamma ray spectrum produced by pair-annihilation into a $b\bar b$ final state versus particle mass.  A $b\bar b$ final state is chosen as a prototypical case for supersymmetric dark matter.  Gamma-ray emission results primarily from the decay of neutral pions among the annihilation products, and the expected gamma-ray spectrum is very similar for other quark-antiquark and gauge boson final states.  Here the spectra are modeled using the DMFIT package \cite{je08} which has been incorporated in the likelihood packages in the Science Tools.  As shown in Fig.~\ref{fig:bblim}, the flux upper limits decrease with increasing particle mass.  For larger particle masses and a $b\bar b$ final state, the expected dark matter annihilation spectrum in the Fermi-LAT energy range becomes significantly harder.  The diffuse backgrounds are relatively soft and at higher energies the LAT PSF and sensitivity improve, leading to stronger limits on models which predict a large number of high energy photons.  The flux limits vary a bit from cluster to cluster as might be expected given variations in the diffuse backgrounds across the sky as well as variations in the proximity of bright gamma-ray sources.

We also consider dark matter annihilating to  a $\mu^+\mu^-$ final state as an example of models giving a good fit to the Pamela anomalous positron fraction data \cite{pamela} and to the Fermi-LAT $e^+e^-$ data \cite{fermiepem}.  (We do not explicitly consider here the case of a $\tau^+\tau^-$ final state, for which we would have obtained intermediate limits between the $\mu^+\mu^-$ and $b\bar b$ cases.)  For a leptonic final state such as $\mu^+\mu^-$, gamma-ray emission results from both internal bremsstrahlung emission from the charged final state, referred to as final state radiation (FSR), as well as from Inverse Compton scattering (IC) of background radiation fields off of the energetic $e^+$ and $e^-$ produced among the annihilation products.  Here we include both sources of emission, but for the IC component, we conservatively only include IC scattering of the CMB.  Scattering of other background radiation fields (such as starlight) could also contribute but are not expected to dominate for these dark matter models.  In clusters, energetic $e^+$ and $e^-$ are expected to lose energy, primarily through IC scattering, much more quickly (energy loss time scales of order $10^{14}$ secs or less for $E_{e^\pm}\gtrsim100$ GeV) than the time it takes for them to diffuse out of the system with any reasonable cosmic ray diffusion setup. The effect of cosmic rays escaping the diffusion region can thus effectively be neglected for clusters, as shown e.g. in Ref.~\cite{Colafrancesco:2005ji}. 

A suppression of the IC signal might occur if synchrotron losses, averaged over the entire volume of the cluster, are comparable or more significant than IC losses. Defining an effective average cluster magnetic field B, this condition amounts to $\bar B\gtrsim B_{\rm cmb}\simeq3.2(1+z)^2\ \mu$G, the latter quantity being the amplitude of a magnetic field having the same energy density as the CMB. While such large magnetic fields are possible, they are unlikely to uniformly populate the cluster volume we consider. If, however, this were the case, the suppression in the IC signal we would get (and thus the relaxation of the constraints we present in Sec.~\ref{sec:results} below) would be on the order of $(B_{\rm cmb}^2/\bar B^2)$.

\begin{figure}[t]
\begin{center}
\includegraphics[width=7.5cm,clip]{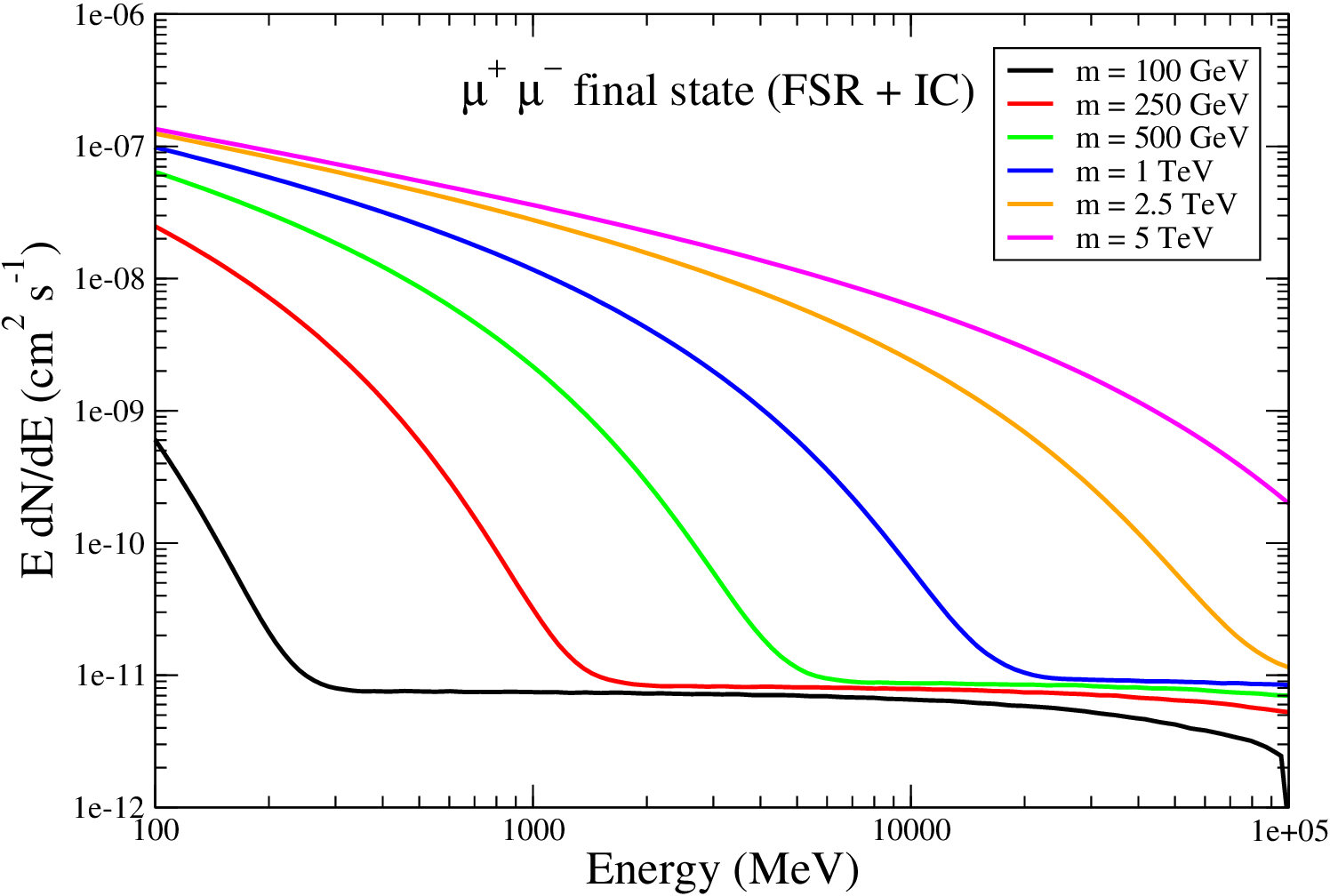}\qquad \includegraphics[width=7.5cm,clip]{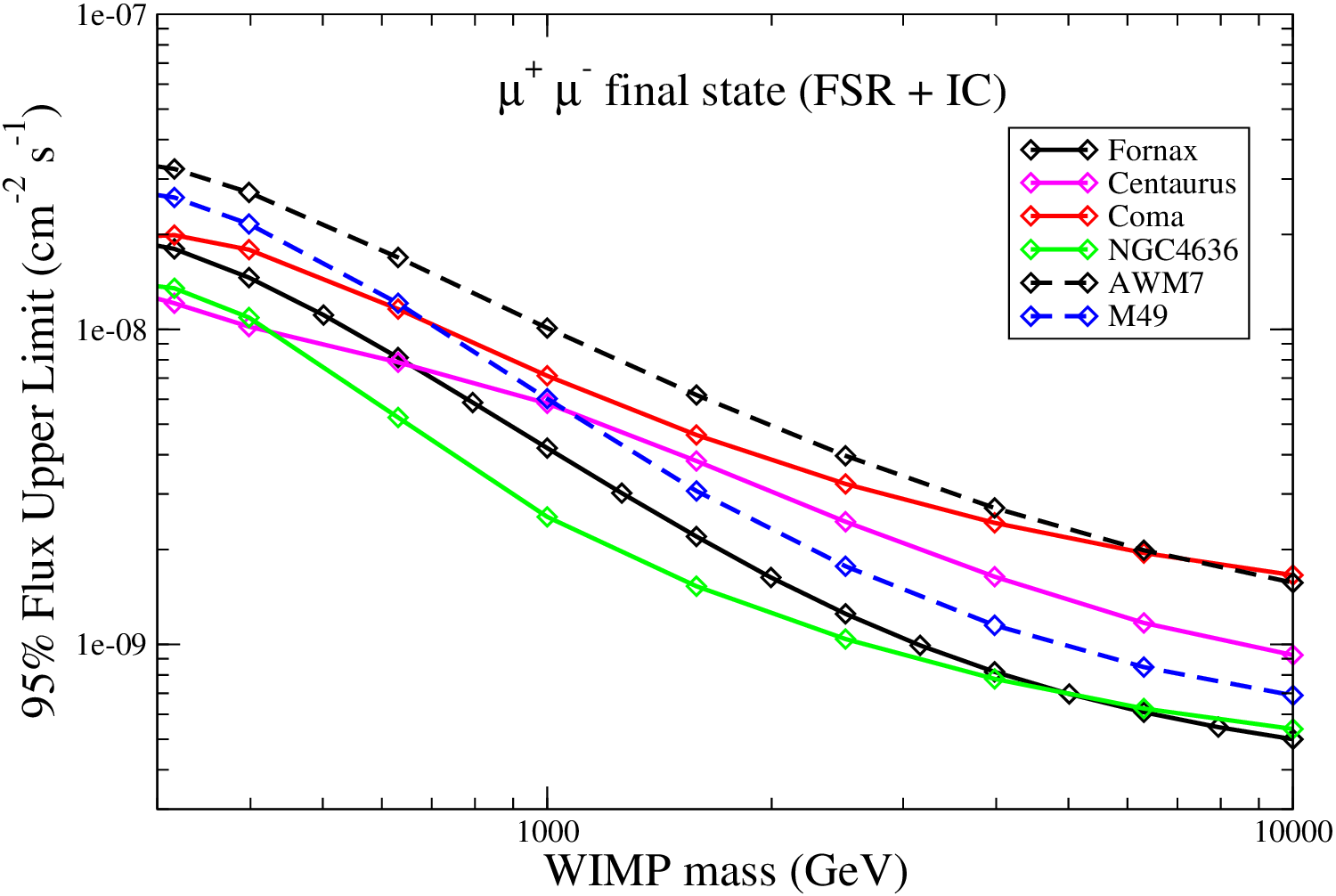}\\
\caption{\textit{Left:} The expected gamma-ray spectrum from dark matter annihilation into a $\mu^+\mu^-$ final state including both FSR and emission from IC scattering of CMB photons for different assumed particle masses.  The spectral normalization is arbitrary, but the same effective $J$ is used in all cases.  \textit{Right:} Flux upper limits as a function of particle mass for an assumed $\mu^+\mu^-$ final state, including the contributions of both FSR and IC gamma-ray emission, for the clusters listed in Table \ref{sample}.  The upper limits reflect the 95\% confidence level limits on the photon flux in the 100 MeV to 100 GeV energy range. \label{fig:mmlim}}
\end{center}
\end{figure}

 In the Fermi-LAT energy range, the IC emission is expected to dominate for dark matter particles annihilating into $\mu^+\mu^-$ for particle masses greater than a couple hundred GeV.  As DMFIT does not include the contribution of IC, we fit the Fermi-LAT data using custom spectra for different assumed particle masses including both the IC and FSR gamma-ray radiation.  Fig.~\ref{fig:mmlim} (left) shows examples of the expected spectra for different particle masses.  Fig.~\ref{fig:mmlim} (right) shows the corresponding upper limits on the gamma-ray flux as a function of particle mass.  For the lowest masses (less than $\sim 200$ GeV) and the highest masses (greater than $\sim 1$ TeV) the expected spectrum is fairly hard; at low masses FSR dominates the emission with a typical $dN/dE \sim E^{-1}$ spectrum where at high masses the IC spectrum itself is fairly hard.  The largest upper limits come for particle masses of a few hundred GeV where the IC emission begins to dominate but the spectrum is fairly soft.

In calculating the upper limits, we have assumed that the clusters are point sources for Fermi.  The gamma-ray signal from dark matter annihilation is proportional to the dark matter density squared making it centrally concentrated.  In the next section, we will describe in detail the modeling of the cluster dark matter density distributions for which we assume an NFW profile.  For the clusters in our sample, which have NFW scale radii between 0.25 and 0.45 degrees, the point source approximation is reasonable.  However, particularly for very hard spectra which predict a significant number of high energy photons where the Fermi-LAT PSF is smaller, some clusters could be slightly extended for Fermi.  We test the effect of modeling clusters with an NFW spatial model compared to a point source model for a hard power law spectrum with an index of 1.  We find that our upper limits are only expected to increase by $10-40$\% when taking in to account the spatial distribution, which is less than the effect of the uncertainty in the cluster density profiles discussed in Sec.~\ref{sec:profile}.  Another possible source of error in the fluxes comes from the systematic uncertainty in the LAT effective area.  This uncertainty is energy dependent and has been estimated to be 10\% at 100 MeV, 5\% at 560 MeV, and 20\% at 10 GeV \cite{Abdo:2009mr}.

\section{Results}\label{sec:results}

\subsection{Constraints on Dark Matter Annihilation Models}\label{sec:dm}


In this section, we explore the constraints on models of dark matter annihilation implied by the non-detection of nearby clusters of galaxies with Fermi-LAT.  The expected gamma-ray flux from dark matter annihilation in a given object depends on both a particle physics factor, for a given annihilation final state or combination of final states the normalization is proportional to the annihilation cross-section divided by the particle mass squared, and an astrophysical factor dependent on the integral of the dark matter density distribution squared along the line of sight, $J$ given in Eq.~1.

Following the results of cosmological collisionless N-body simulations, we assume a Navarro-Frenk-White profile \cite{nfw} for the dark matter density distribution $\rho(r)$ of the clusters in our sample, where $r$ is the radial distance from the center of the cluster (which we assume to be spherically symmetric):
\begin{equation}
\rho_{\rm NFW}(r)=\frac{\rho_s}{(r/r_s)\left(1+r/r_s\right)^2}.
\end{equation}
We calculate the values of the scaling density $\rho_s$ and scaling radius $r_s$ from the cluster virial mass, as determined from X-ray observations in Ref.~\cite{re02}, and from the observationally determined concentration-mass relation of Ref.~\cite{conc}:
\begin{equation}
c_{\rm vir}(M)=9\times\left(\frac{M_{\rm vir}}{10^{14}M_\odot h^{-1}}\right)^{-0.172}
\end{equation}
where the virial concentration $c_{\rm vir} = R_{\rm vir}/r_s$ for an NFW profile.  The $J$ values, within an angular region of one degree radius, determined by this method are listed in Table \ref{sample}.  The errors on $J$ listed in Table \ref{sample} reflect the uncertainties in the X-ray virial mass determinations \cite{re02}, but below we will also discuss alternate determinations of the density profiles for a couple of the best candidate clusters.

\begin{figure}[t]
\begin{center}
\mbox{\includegraphics[width=7.5cm,clip]{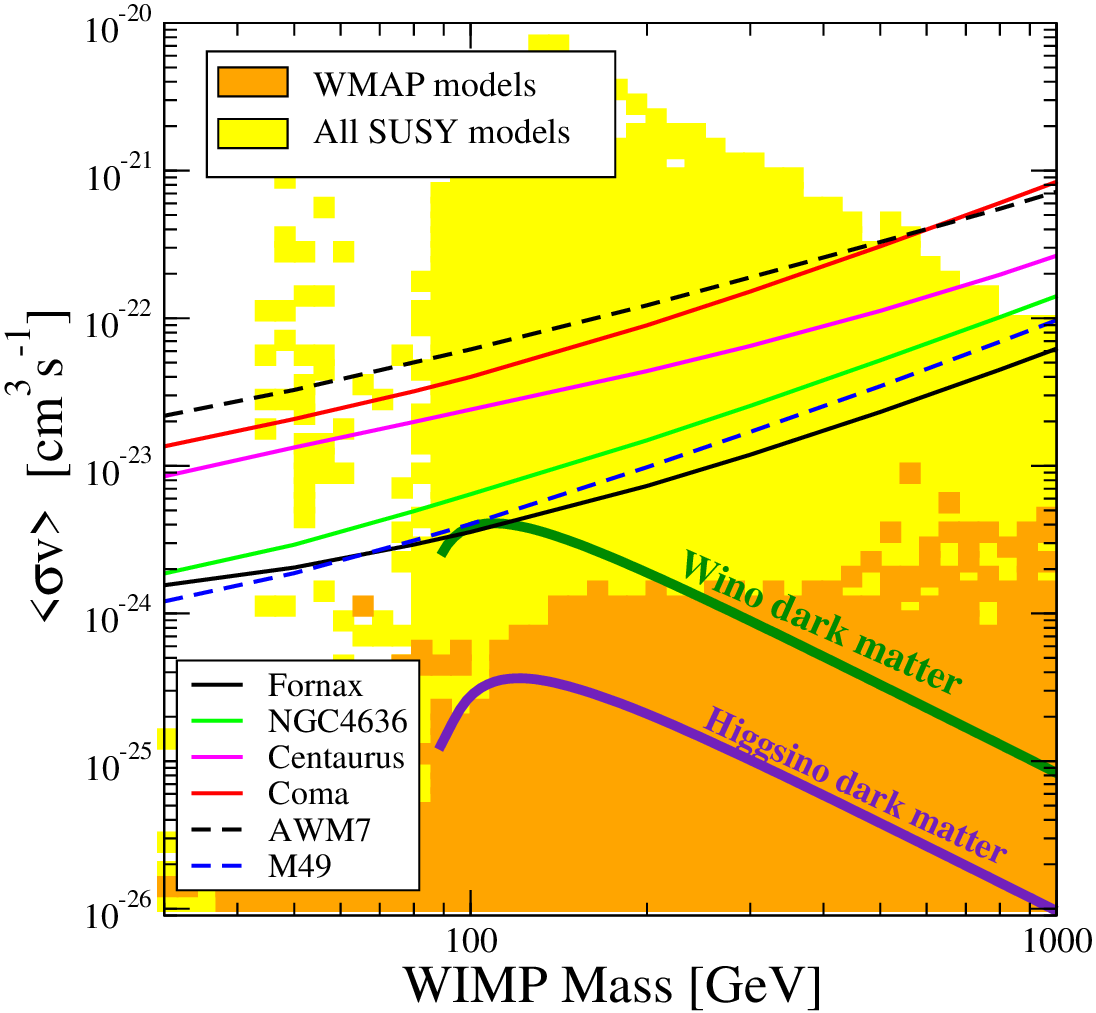}\qquad \includegraphics[width=7.5cm,clip]{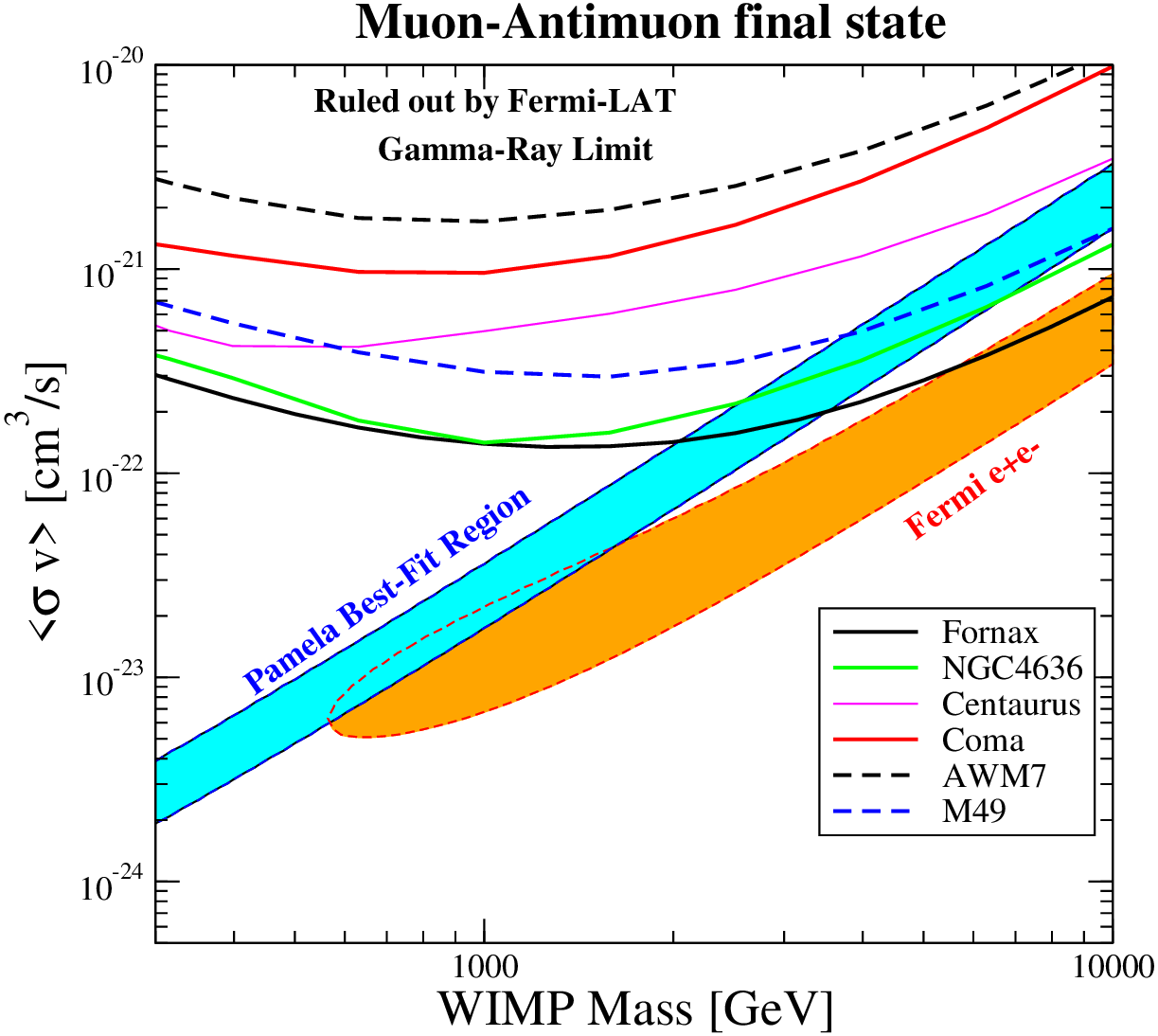}}\\
\caption{Upper limits on the annihilation cross-section for a $b\bar b$ final state (left panel) and a $\mu^+\mu^-$ final state (right panel) as a function of WIMP mass for all clusters in the sample.  In the left panel, the constraints are compared to the predicted cross section for generic SUSY WIMP models (yellow) in the general MSSM \cite{Profumo:2005xd} as well as to the subset of models that have thermal relic densities compatible with the observed universal dark matter density (orange) \cite{wmap5}.   In the right panel, the cluster gamma-ray constraints are compared to values of the mass and annihilation cross section for dark matter models annihilating to a $\mu^+\mu^-$ final state which provide good fits to the Pamela (blue) or Fermi-LAT (orange) $e^+e^-$ data \cite{Grasso:2009ma}.  \label{fig:dmlim}}
\end{center}
\end{figure}

Fig.~\ref{fig:dmlim} shows the upper limits on the annihilation cross-section for a $b\bar b$ final state (left panel) and a $\mu^+\mu^-$ final state (right panel) as a function of WIMP mass for all clusters given the upper limits on their gamma-ray fluxes (Fig.~\ref{fig:bblim} and Fig.~\ref{fig:mmlim}) and the modeling of their dark matter density profiles.  The tightest constraints are given by nearby groups of galaxies, Fornax, M49, and NGC 4636, which have the largest $J$ values.  For the case of a $b\bar b$ final state with no contribution from substructure, the cluster constraints exclude only models with relatively high annihilations cross-sections (which would produce low thermal relic densities) and are generally weaker than those found for nearby dwarf spheroidal galaxies (dSph) \cite{dsph}.

For a $\mu^+\mu^-$ final state, models fitting the Pamela positron excess with masses above $2-3$ TeV are ruled out by the non-detection of the Fornax group.  Similar constraints were derived using the Ursa Minor dSph galaxy \cite{dsph} for a diffusion coefficient of $D=10^{28}$ cm$^2$/s.  For a larger effective diffusion coefficient, cosmic rays would leave the dwarf system at much larger rates, and the IC emission would be significantly reduced, giving weaker constraints for dSphs compared to clusters.  As argued in Sec.~2.3, the constraints shown here for clusters are therefore more robust in that the cosmic ray escape rate for high-energy $e^+$ and $e^-$ is virtually negligible for cluster size systems.

Our results improve significantly the constraints on particle dark matter obtained from observations of nearby clusters of galaxies by Atmospheric Cherenkov Telescopes, namely observations of the Coma cluster by H.E.S.S. \cite{hesscoma} and of the Perseus cluster by MAGIC \cite{magicperseus}. Notice that we do not consider here the search for monochromatic gamma-ray lines, since the predicted constraints from observations of galaxy clusters are much weaker than those obtained in the dedicated search of Ref.~\cite{fermilatlines}.

\subsection{Effects of Substructure}\label{sec:substr}

The above constraints all assumed only the smooth cluster halo contribution to the dark matter density.  However, N-body simulations in a CDM cosmology show a large spectrum of dark matter substructures in cluster and galaxy halos \cite{Diemand:2008in,Springel:2008cc}.  For typical thermally produced WIMP models, these substructures are expected to exist down to cutoff scales of between $10^{-6}$ and $10^2$ Earth masses \cite{Profumo:2006bv,Bringmann:2006mu,Bringmann:2009vf}.   These dense dark matter substructures can add significantly to the gamma-ray flux from annihilation.  At the very least, massive groups and clusters are known to host hundreds to thousands of galaxies ranging in size from giant ellipticals down to dwarf galaxies.

We estimate the impact of substructures on $J$ (the integral in Eq.~(\ref{eq:jpsi})) and the corresponding boost in the expected gamma-ray flux from clusters in our sample using the following procedure.  First, we assume the concentration mass relation for small scale halos found in the numerical simulations of Ref.~\cite{2008ApJ...686..262K}, which contains a radial dependence accounting for the tidal stripping of substructure close to the center of the host halo. We assume the radial scale of said dependence to be twice the virial radius of the host halo. We also verified that neglecting the radial dependence of the concentration mass relation for subhalos only affects the resulting substructure enhancement by less than 20\%. We assume a spectral index of $-1.9$ for the subhalo mass function, and we consider two realizations for the substructure distribution:
\begin{itemize}
\item A {\em conservative} setup where the fraction $f_s$ of cluster mass in substructure equals 10\%, and where we assume no substructure (i.e.~we effectively cut-off the integration on the subhalo mass function) below the mass scale of a dwarf galaxy dark matter halo, $M_{\rm cut}=10^7\ M_\odot$.  We consider this setup to be conservative as we know that clusters at least contain galaxies.
\item An {\em optimistic} setup, where $f_s=0.2$ and $M_{\rm cut}=10^{-6}\ M_\odot$ \cite{Green:2005fa} (even smaller cutoff scales are however possible in well-motivated WIMP scenarios \cite{Profumo:2006bv} -- see also \cite{Bringmann:2009vf} for a review).
\end{itemize}
These substructure setups in terms of the mass function of subhalos, the mass fraction in subhalos, and the subhalo mass-concentration relation were chosen based on the results of recent N-body simulations \cite{Springel:2008cc, Diemand:2008in, Gao:2004au}.  We verified that the substructure setups we consider here are consistent with the alternative framework presented in Ref.~\cite{Colafrancesco:2005ji}. Specifically, the optimistic setup corresponds to a substructure contrast factor $\Delta^2$ (in the notation of Ref.~\cite{Colafrancesco:2005ji}) of $\Delta^2\simeq3\times 10^5$, while the conservative setup to one with $\Delta^2\simeq1.3\times 10^5$, for the given specified substructure mass fractions.

\begin{figure}[t]
\begin{center}
\mbox{\includegraphics[width=7.5cm,clip]{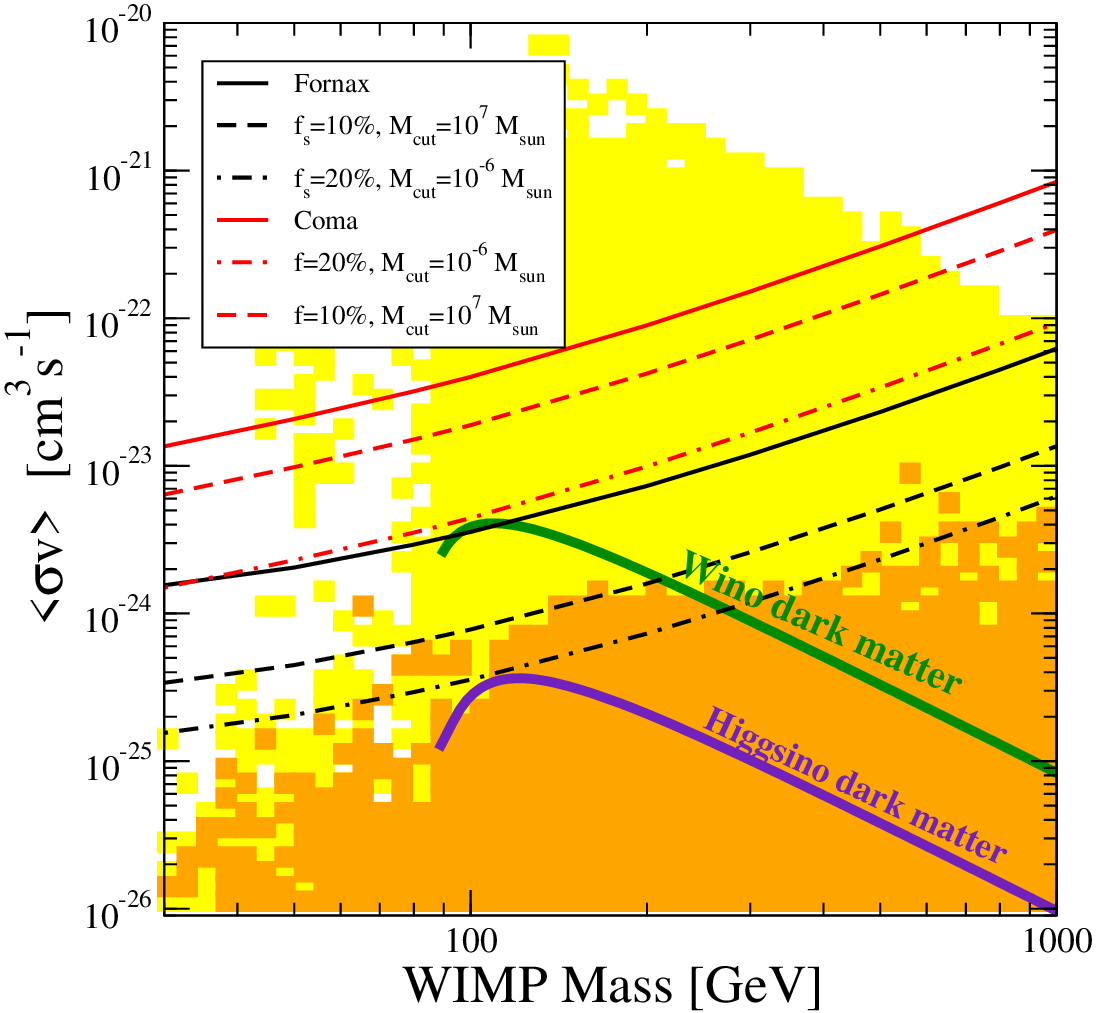}\qquad \includegraphics[width=7.5cm,clip]{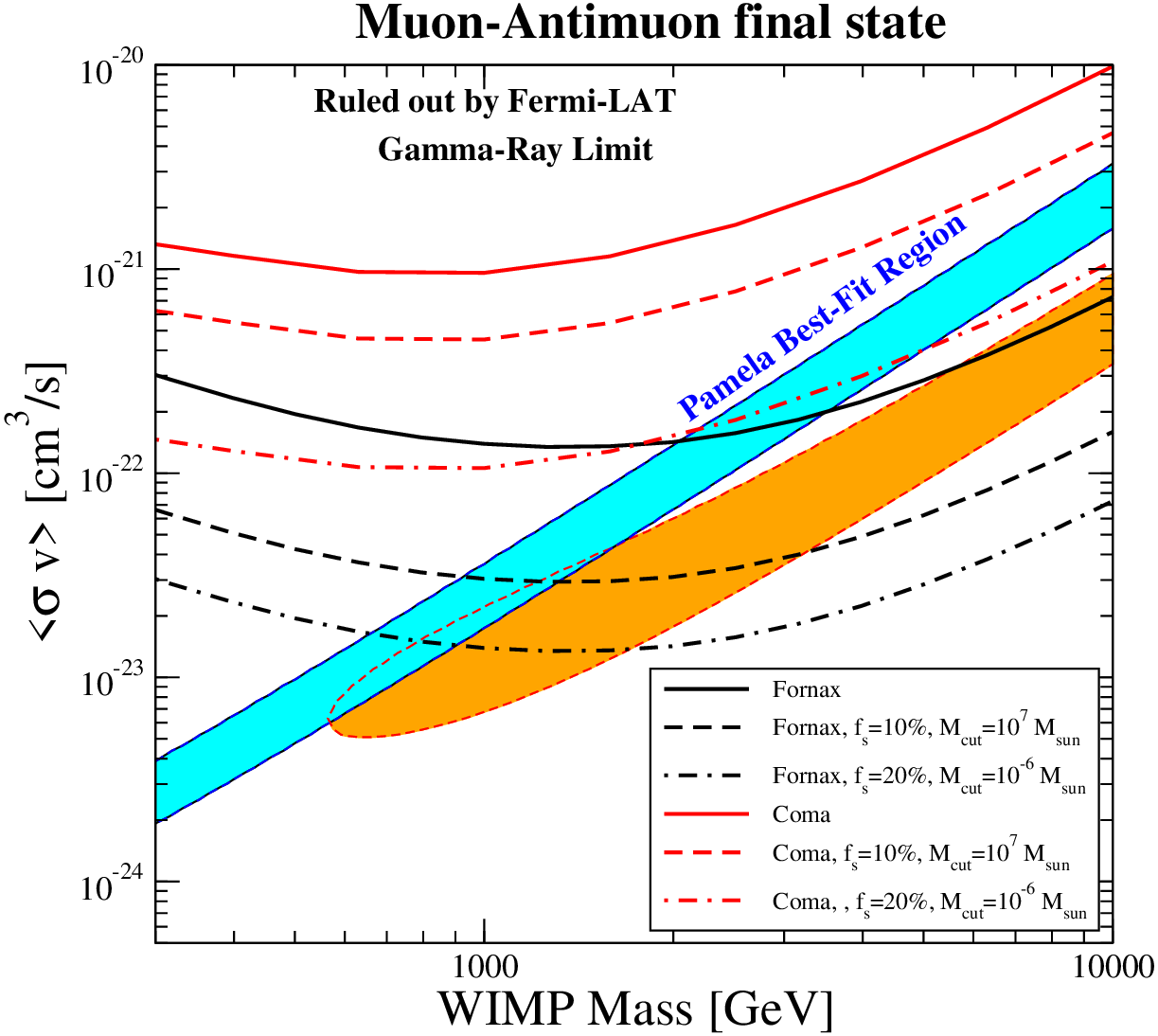}}\\
\caption{Upper limits on the annihilation cross-section for a $b\bar b$ final state (left panel) and a $\mu^+\mu^-$ final state (right panel) for the Coma and Fornax clusters including the effect of substructure on the expected gamma-ray signal.  The constraints are shown for no substructure (solid lines), our conservative substructure setup which includes only substructure of dwarf galaxy mass or larger (dashed lines), and our optimistic setup which includes substructure down to $M_{\rm cut}=10^{-6}\ M_\odot$ (dot-dashed lines).  The dark matter models are the same as in Fig.~\ref{fig:dmlim}.  \label{fig:dmsubs}}
\end{center}
\end{figure}

Fig.~\ref{fig:dmsubs} shows the effect of adding substructure to the dark matter constraints shown in Fig.~\ref{fig:dmlim} for two example clusters, Fornax (which has the highest $J$ for a smooth halo) and Coma (the most massive cluster in the sample).  As in the previous section, we show the constraints for a pure $b\bar b$ final state (left panel) and a $\mu^+\mu^-$ final state (right panel).  The dashed lines show the results for our conservative substructure setup which includes only the expected contribution from galactic scale substructure and gives a boost to the expected gamma-ray flux of $\sim4.6$ for Fornax and $\sim2.1$ for Coma.  In this setup, the Fornax constraints exclude models fitting the Pamela $e^+e^-$ data with masses above $1$ TeV for a $\mu^+\mu^-$ final state and begin to probe thermally produced WIMP models with a relic density consistent with the observed universal matter density for a $b\bar b$ final state.  

In the more optimistic substructure setup (dot-dashed lines in Fig.~\ref{fig:dmsubs}), where we include substructure down to roughly the expected substructure cutoff scale for WIMP models, the predicted boosts to the cluster gamma-ray flux are even higher, $\sim10$ for Fornax and $\sim9$ for Coma. In this case, the left panel indicates that winos lighter than 200 GeV are ruled out, as is a dark matter interpretation of the Pamela positron fraction with a $\mu^+\mu^-$ final state and a sub-TeV mass.  In general, models with a $\mu^+\mu^-$ final state with cross sections greater than $\sim 10^{-23}$ cm$^3$ s$^{-1}$ and particle masses greater than $\sim 1$ TeV are excluded which includes most of the parameter space fitting the Fermi-LAT $e^+e^-$ data.

Similar to our conclusions here, Ref.~\cite{Pinzke:2009cp} predicts that for a particle mass of 1.6 TeV with a cross-section of $3 \times 10^{-23}$ cm$^3$ s$^{-1}$ annihilating to a pure $\mu^+\mu^-$ final state, Fermi-LAT should detect local clusters if the cut-off scale for substructures is $10^4 M_\odot$ or less.

\subsection{Uncertainty in the Density Profiles}\label{sec:profile}

In this section, we consider alternative determinations of cluster dark matter density profiles.  Our primary results are based on cluster dark matter density profiles from X-ray observations, and Table \ref{sample} lists the uncertainties in $J$ stemming from the X-ray virial mass determinations \cite{re02} only, typically about 10\% in $J$.  However, many of the clusters in the sample have masses and density profiles determined through other observables, and these perhaps better reflect the total uncertainties.  For example, the density distribution of the Coma cluster has been determined through both weak lensing \cite{Gavazzi:2009kf, Kubo:2007wt} and galaxy dynamics \cite{Lokas:2003ks}.  Based on the best-fit NFW profiles given in these works, the weak lensing measurements in \cite{Kubo:2007wt} based on SDSS imaging and galaxy dynamics from \cite{Lokas:2003ks} give $J$ values a factor of 2 higher while the weak lensing measurements from \cite{Gavazzi:2009kf} based on CFHT imaging give $J$ a factor of two lower than the value quoted in Table \ref{sample}.

\begin{figure}[t]
\begin{center}
\mbox{\includegraphics[width=7.5cm,clip]{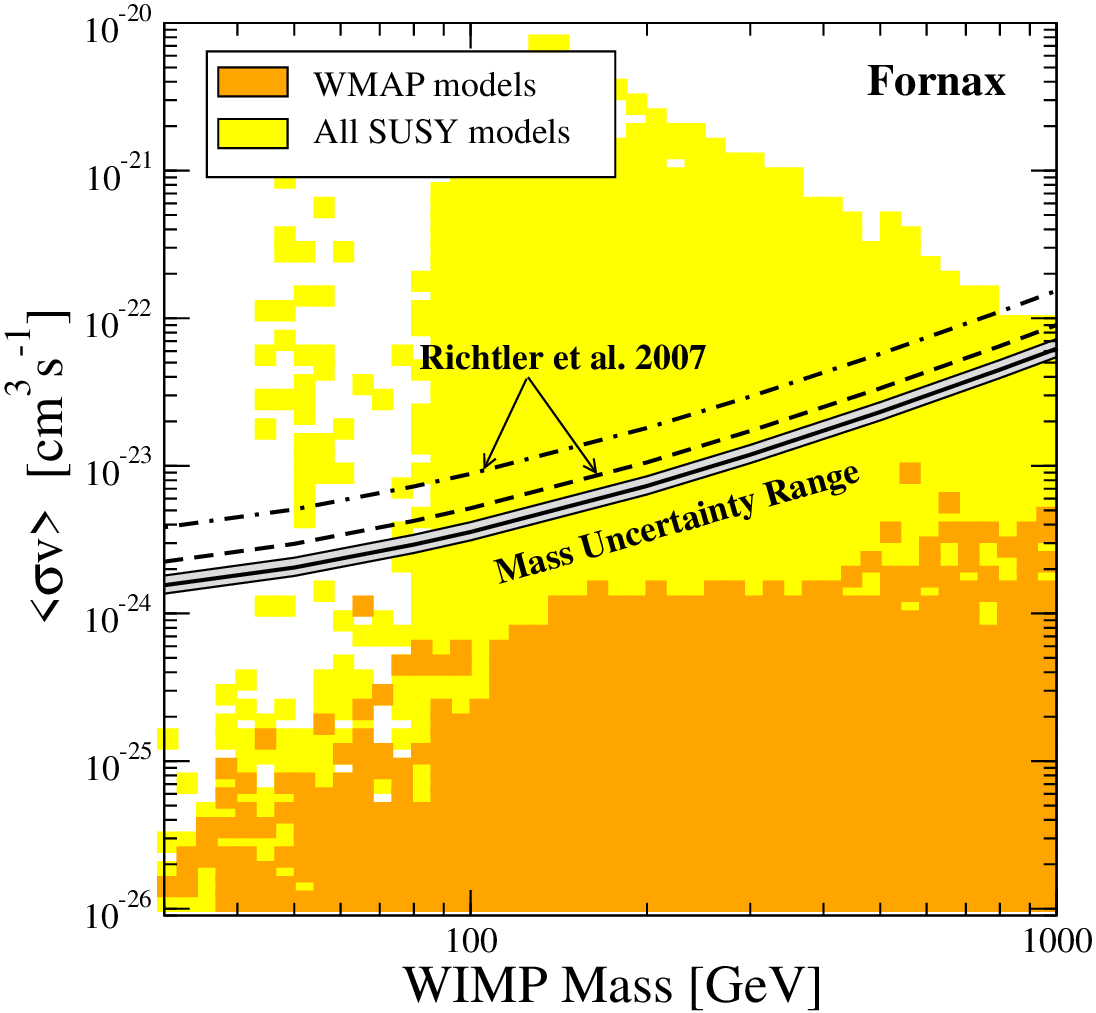}\qquad \includegraphics[width=7.5cm,clip]{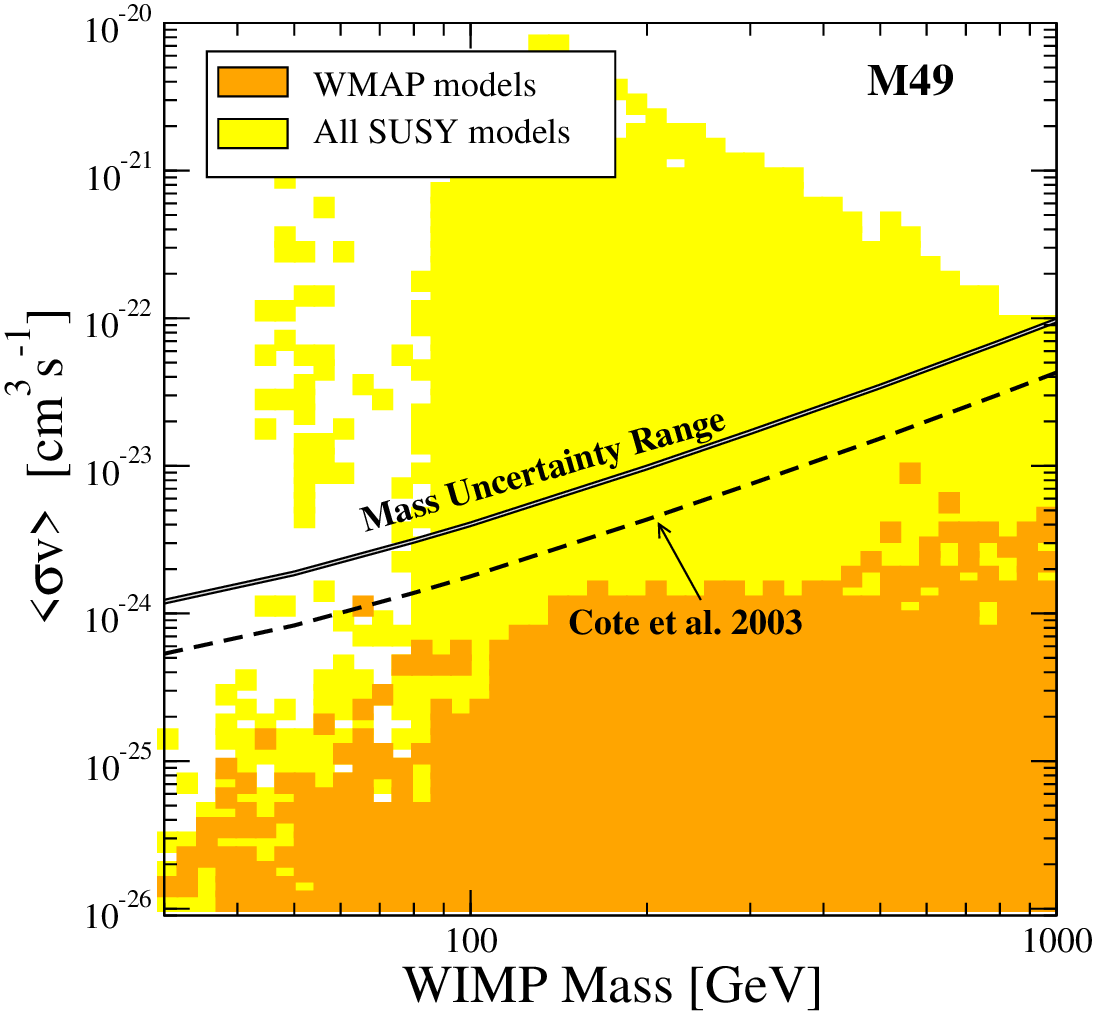}}\\
\caption{Comparison of the annihilation cross-section constraints based on the X-ray modeling of the cluster density profiles described in Section \ref{sec:dm} to the constraints using other observables for the two systems with the highest $J$, Fornax (left panel) and M49 (right panel).  The solid lines and grey bands show the reference constraints and errors for $J$ from Table \ref{sample}.  For Fornax, these constraints are compared to modeling of the dark matter profile from globular cluster dynamics presented in \cite{Richtler:2007jd} for both their full (dashed line) and "safe" (dot-dashed) globular cluster samples.  For M49, we compare to the density profile determined by \cite{Cote:2003er} through fitting a combination of stellar kinematic and X-ray data.  In all cases, the dark matter density profile is parameterized with an NFW profile.  A $b\bar b$ final state is assumed, and the dark matter models are the same as in the left panel of Fig.~\ref{fig:dmlim}. \label{fig:dmprof}}
\end{center}
\end{figure}

Based on the X-ray modeling, the nearby Fornax and M49 groups have the highest $J$ values leading to the tightest constraints on dark matter models, so we specifically consider alternate determinations of the dark matter densities of these systems.  The density profile of the central region of Fornax has been determined based on observations of the dynamics of globular clusters in its dominant central galaxy \cite{Richtler:2007jd}.  Fig.~\ref{fig:dmprof} (left panel) compares the constraints on WIMP models for $J$ as quoted in Table \ref{sample} for the smooth halo only, including the errors, to the corresponding constraints using $J$ as determined from the best-fit NFW profile from the globular cluster sample in \cite{Richtler:2007jd}.  Here the dashed line shows the fit quoted in \cite{Richtler:2007jd} for the full globular cluster sample while the dot-dashed line shows the fit to their ``safe'' subsample which employed a more conservative selection to reduce outliers; for these profiles, the expected gamma-ray flux from Fornax is reduced slightly by factors of 1.4 and 2.4, respectively.  In Fig.~\ref{fig:dmprof}, the comparison is only shown for a $b\bar b$ final state, but the same shift in normalization is expected for all annihilation final states.

A determination of the dark matter density profile of M49 based on a combination of stellar kinematic and X-ray data is presented in \cite{Cote:2003er} and compares quite well to the globular cluster kinematics presented in the same work.  Fig.~\ref{fig:dmprof} (right panel) shows the comparison of the constraints for this profile.  The density profile of  \cite{Cote:2003er} implies an enhancement of the gamma-ray flux relative to what is predicted from Table \ref{sample} by a factor of 2.3.

In summary, determinations of $J$ for some of the best candidate clusters in our sample based on different observables vary by roughly a factor of two, and the X-ray measurements we employ do not yield $J$ values systematically higher or lower than other methods.  In all cases, we have considered only the NFW functional form for the dark matter mass profile.  However, since the intergral in Eq.~(\ref{eq:jpsi}) is taken over a large volume and for a very distant object, the assumed inner slope of the density profile does not have a large effect on the predicted gamma-ray flux.  Alternate density profiles such as the cored Burkert profile \cite{Burkert:1995yz} or the centrally steeper Moore profile \cite{Moore98} give predicted fluxes about 40\% lower (Burkert) or higher (Moore) than the NFW profile, similar to or smaller than the differences in flux presented above stemming from different observational determinations of the density profiles.

\section{Summary}

Clusters of galaxies are the most massive collapsed objects in the Universe and are very dark matter dominated, making them potentially bright sources of gamma-ray emission from dark matter annihilation.  No significant  gamma-ray emission has been detected from clusters of galaxies in the first 11 months of Fermi-LAT survey mode observations \cite{keith}.  In this paper, we explored the implications of the non-detection of clusters by Fermi-LAT in terms of constraints on models of dark matter annihilation.  In particular, we focused on the six best candidate clusters and groups of galaxies in the context of searches for gamma-ray emission from dark matter pair annihilation \cite{je09}  after excluding from the sample 
clusters which host bright central AGN or lie close to the Galactic plane.  We analyzed the Fermi-LAT data to derive upper limits on the gamma-ray flux from dark matter annihilation in specific models, self-consistently incorporating the expected spectral shape for a given particle mass and annihilation final state.  We conservatively assume only gamma-ray emission from dark matter annihilation when interpreting the upper limits, though cosmic rays generated through astrophysical processes like shocks could also lead to gamma-rays from clusters.  A much larger sample of clusters and a more general discussion of clusters as potential gamma-ray sources will be presented in a forthcoming paper.

In particular, so-called lepto-philic dark matter models with relatively high particle masses and annihilation cross-sections proposed to fit the Pamela and Fermi-LAT $e^+ e^-$ data would lead to significant IC gamma-ray emission in clusters \cite{Pinzke:2009cp}.   Here we conservatively only consider IC scattering of the CMB.  Including the contribution of substructure in clusters on galactic scales, known to exist, the non-detection of local clusters and groups excludes models fitting the Pamela positron excess with masses above $1$ TeV for a $\mu^+\mu^-$ final state.  Even for a smooth NFW dark matter profile with no substructure, models fitting the Pamela data with masses $> 2-3$ TeV are excluded.  Unlike similar constraints placed by the non-detection of dwarf spheroidal galaxies by Fermi-LAT \cite{dsph} which depend sensitively on the model employed for the spatial diffusion of $e^+ e^-$, the leakage of cosmic rays due to spatial diffusion in clusters is not expected to be significant or significantly reduce the IC gamma-ray emission.

With the inclusion of galactic scale substructure, the cluster gamma-ray constraints are beginning to probe some dark matter models annihilating to a $b\bar b$ or similar final state -- producing a gamma-ray spectrum similar to what is expected in popular particle dark matter models, such as the neutralino -- with thermal relic densities consistent with the observed universal matter density.  Including the contribution from smaller mass dark matter substructures, the constraints we obtain become increasingly tighter.  In general, however, these constraints are a bit weaker than (or in the most optimistic case similar to) the constraints derived from the non-detection by Fermi-LAT of the best candidate dwarf spheroidal galaxies \cite{dsph}.

Clusters of galaxies, an as-yet gamma-ray quiet source class, will continue to be monitored as the Fermi Telescope increases its observing time in survey mode. In addition, future investigations will improve these constraints through stacking of clusters, improvements in our understanding of the gamma-ray backgrounds, and searches for extended emission around clusters like Perseus which host central gamma-ray AGN.  While the detection of a gamma-ray signal from clusters might be attributed to a variety of source mechanisms, including cosmic rays from merger or accretion shocks and particle dark matter annihilation or decay, as shown in the present study even null results imply interesting and compelling constraints.


  
\section*{Acknowledgments}

The \emph{Fermi}/LAT Collaboration acknowledges generous ongoing
  support from a number of agencies and institutes that have supported
  both the development and the operation of the LAT as well as
  scientific data analysis.
These include the National Aeronautics and Space Administration and
  the Department of Energy in the United States, the Commissariat \`a
  l'Energie Atomique and the Centre National de la Recherche
  Scientifique / Institut National de Physique Nucl\'eaire et de
  Physique des Particules in France, the Agenzia Spaziale Italiana and
  the Istituto Nazionale di Fisica Nucleare in Italy, the Ministry of
  Education, Culture, Sports, Science and Technology (MEXT), High
  Energy Accelerator Research Organization (KEK) and Japan Aerospace
  Exploration Agency (JAXA) in Japan, and the K.~A.~Wallenberg
  Foundation, the Swedish Research Council and the Swedish National
  Space Board in Sweden. Additional support for science analysis during the operations phase is gratefully acknowledged from the Istituto Nazionale di Astrofisica in Italy and the Centre National d'\'Etudes Spatiales in France.




\newpage
\clearpage
\section*{References}

\begin{thebibliography}{200}

\bibitem{Colafrancesco:2005ji}
  S.~Colafrancesco, S.~Profumo and P.~Ullio,
  Astron.\ Astrophys.\  {\bf 455}, 21 (2006)
  [arXiv:astro-ph/0507575].
  
\bibitem{je09} 
T.~E.~Jeltema, J.~Kehayias \& S.~Profumo, 
Phys.\ Rev.\  D {\bf 80}, 023005 (2009).

\bibitem{Pinzke:2009cp}
  A.~Pinzke, C.~Pfrommer and L.~Bergstrom,
  Phys.\ Rev.\ Lett.\  {\bf 103}, 181302 (2009)
  arXiv:0905.1948 [astro-ph.HE].

  \bibitem{pamela}
 O.~Adriani {\it et al.}  [PAMELA Collaboration],
  Nature {\bf 458}, 607 (2009)
  [arXiv:0810.4995 [astro-ph]].
  
\bibitem{fermiepem}
   A.~A.~Abdo {\it et al.}  [Fermi-LAT Collaboration],
  Phys.\ Rev.\ Lett.\  {\bf 102} (2009) 181101
  [arXiv:0905.0025 [astro-ph.HE]].

\bibitem{hessepem}
  F.~Aharonian {\it et al.}  [H.E.S.S. Collaboration],
  Phys.\ Rev.\ Lett.\  {\bf 101}, 261104 (2008)
  [arXiv:0811.3894 [astro-ph]].

\bibitem{Grasso:2009ma}
  D.~Grasso {\it et al.},
  Astropart.\ Phys.\  {\bf 32}, 140 (2009)
  arXiv:0905.0636 [astro-ph.HE].

\bibitem{Bergstrom:2009fa}
  L.~Bergstrom, J.~Edsjo and G.~Zaharijas,
  Phys.\ Rev.\ Lett.\  {\bf 103}, 031103 (2009)
  [arXiv:0905.0333 [astro-ph.HE]].
  
\bibitem{Meade:2009iu}
  P.~Meade, M.~Papucci, A.~Strumia and T.~Volansky,
  Nucl.\ Phys.\  B {\bf 831}, 178 (2010)
  [arXiv:0905.0480 [hep-ph]].  
  
\bibitem{keith} K.~Bechtol, talk at TeV Particle Astrophysics 2009, http://www-conf.slac.stanford.edu/tevpa09/Talks.asp, (2009).


\bibitem{n1275} A.~A.~Abdo, et al.  [Fermi-LAT Collaboration], Astrophys.\ J.\ {\bf 699}, 31 (2009)

\bibitem{m87}
  A.~A.~Abdo {\it et al.}  [Fermi-LAT Collaboration],
  Astrophys.\ J.\  {\bf 707}, 55 (2009)
  [arXiv:0910.3565 [astro-ph.HE]].

\bibitem{re02}
  T.~H.~Reiprich and H.~Boehringer,
  Astrophys.\ J.\  {\bf 567}, 716 (2002)

\bibitem{Chen:2007sz}
  Y.~Chen, T.~H.~Reiprich, H.~Bohringer, Y.~Ikebe and Y.~Y.~Zhang,
  Astron.\ Astrophys.\  {\bf 466}, 805 (2007)
  arXiv:astro-ph/0702482.

\bibitem{dsph} A.~A.~Abdo, et al. [Fermi-LAT Collaboration], Astrophys.\ J., accepted (2009) [arXiv:1001.4531].

\bibitem{LAT}
  W.~B.~Atwood {\it et al.}  [Fermi-LAT Collaboration],
  Astrophys.\ J.\  {\bf 697}, 1071 (2009)

\bibitem{ca79} W.~Cash, Astroph. J., {\bf 228}, 939 (1979).

\bibitem{ma96} J.~R.~Mattox, et al., Astroph. J., {\bf 461}, 396 (1996).

\bibitem{1fgl}
  A.~A.~Abdo {\it et al.}  [Fermi-LAT Collaboration],
  Astrophys.\ J.\ Supplements, submitted, (2010).

\bibitem{je08}
  T.~E.~Jeltema and S.~Profumo,
  JCAP {\bf 0811}, 003 (2008)
  [arXiv:0808.2641 [astro-ph]].

\bibitem{jeltemaprofumodwarfs}
T.~E.~Jeltema, \& S.~Profumo, S., Astroph. J., {\bf 686}, 1045 (2008).

\bibitem{Abdo:2009mr}
  A.~A.~Abdo {\it et al.}  [Fermi-LAT Collaboration],
  Phys.\ Rev.\ Lett.\  {\bf 103}, 251101 (2009)
  [arXiv:0912.0973 [astro-ph.HE]].

\bibitem{nfw}
  J.~F.~Navarro, C.~S.~Frenk and S.~D.~M.~White,
  Astrophys.\ J.\  {\bf 490}, 493 (1997)
  [arXiv:astro-ph/9611107].

\bibitem{conc}
  D.~A.~Buote, F.~Gastaldello, P.~J.~Humphrey, L.~Zappacosta, J.~S.~Bullock, F.~Brighenti and W.~G.~Mathews,
  Astrophys.\ J.\  {\bf 664}, 123 (2007)
  [arXiv:astro-ph/0610135].

\bibitem{Profumo:2005xd}
  S.~Profumo,
  Phys.\ Rev.\  D {\bf 72}, 103521 (2005)
  [arXiv:astro-ph/0508628].

\bibitem{wmap5}
  E.~Komatsu {\it et al.}  [WMAP Collaboration],
  Astrophys.\ J.\ Suppl.\  {\bf 180}, 330 (2009)
  [arXiv:0803.0547 [astro-ph]].

  \bibitem{hesscoma}
   F.~A.~Aharonian  {\it et al.} [H.E.S.S. Collaboration],
  Astron.\ Astrophys.\  {\bf 502}, 437 (2009)
  arXiv:0907.0727 [astro-ph.CO].

\bibitem{magicperseus}
  J.~Aleksic {\it et al.}  [MAGIC Collaboration],
  Astrophys.\ J.\  {\bf 710}, 634 (2010)
  arXiv:0909.3267 [astro-ph.HE].

\bibitem{fermilatlines}
A.~A.~Abdo, M.~Ackermann and M.~Ajello,
  arXiv:1001.4836 [astro-ph.HE].
 
\bibitem{Diemand:2008in}
  J.~Diemand, M.~Kuhlen, P.~Madau, M.~Zemp, B.~Moore, D.~Potter and J.~Stadel,
  Nature {\bf 454}, 735 (2008)
  [arXiv:0805.1244 [astro-ph]].
  
\bibitem{Springel:2008cc}
  V.~Springel {\it et al.},
  Mon.\ Not.\ Roy.\ Astron.\ Soc.\  {\bf 391}, 1685 (2008)
  [arXiv:0809.0898 [astro-ph]].

\bibitem{Profumo:2006bv}
  S.~Profumo, K.~Sigurdson and M.~Kamionkowski,
  Phys.\ Rev.\ Lett.\  {\bf 97} (2006) 031301
  [arXiv:astro-ph/0603373].

\bibitem{Bringmann:2006mu}
  T.~Bringmann and S.~Hofmann,
  JCAP {\bf 0407} (2007) 016
  [arXiv:hep-ph/0612238].

\bibitem{Bringmann:2009vf}
  T.~Bringmann,
  New J.\ Phys.\  {\bf 11} (2009) 105027
  [arXiv:0903.0189 [astro-ph.CO]].

\bibitem{2008ApJ...686..262K} 
M.~Kuhlen, J.~Diemand, \& P.~Madau, 
Astroph. J., {\bf 686}, 262 (2008).

\bibitem{Green:2005fa}
  A.~M.~Green, S.~Hofmann and D.~J.~Schwarz,
  JCAP {\bf 0508} (2005) 003
  [arXiv:astro-ph/0503387].

\bibitem{Gao:2004au}
  L.~Gao, S.~D.~M.~White, A.~Jenkins, F.~Stoehr and V.~Springel,
  Mon.\ Not.\ Roy.\ Astron.\ Soc.\  {\bf 355}, 819 (2004)
  [arXiv:astro-ph/0404589].

\bibitem{Gavazzi:2009kf}
  R.~Gavazzi {\it et al.},
  Astron.\ Astrophys.\  {\bf 498}, L33 (2009)
  arXiv:0904.0220 [astro-ph.CO].

\bibitem{Kubo:2007wt}
  J.~M.~Kubo, A.~Stebbins, J.~Annis, I.~P.~Dell'Antonio, H.~Lin, H.~Khiabanian and J.~A.~Frieman,
  Astrophys.\ J.\  {\bf 671}, 1466 (2008)
  [arXiv:0709.0506 [astro-ph]].

\bibitem{Lokas:2003ks}
  E.~L.~Lokas and G.~A.~Mamon,
  Mon.\ Not.\ Roy.\ Astron.\ Soc.\  {\bf 343}, 401 (2003)
  [arXiv:astro-ph/0302461].

\bibitem{Richtler:2007jd}
  T.~Richtler, Y.~Schuberth, M.~Hilker, B.~Dirsch, L.~Bassino and A.~J.~Romanowsky,
  Astron.\ Astrophys.\  {\bf 478}, L23 (2008)
  arXiv:0711.4077 [astro-ph].

\bibitem{Cote:2003er}
  P.~Cote, D.~E.~McLaughlin, J.~G.~Cohen and J.~P.~Blakeslee,
  Astrophys.\ J.\  {\bf 591}, 850 (2003)
  [arXiv:astro-ph/0303229].

\bibitem{Burkert:1995yz}
  A.~Burkert,
  Astrophys.\ J.\  {\bf 447}, L25 (1995)
  [arXiv:astro-ph/9504041].

\bibitem{Moore98} 
B.~Moore, F.~Governato, T.~Quinn, J.~Stadel, \& G.~Lake,
Astrophys.\ J.\  {\bf 499}, L5 (1998).

\end{thebibliography}
\end{document}